# Discovery of Maximal Frequent Item Sets using Subset Creation

Jnanamurthy HK, Vishesh HV, Vishruth Jain, Preetham Kumar, Radhika M. Pai

Department of Information and Communication Technology
Manipal Institute of Technology, Manipal University, Manipal-576104, India
jnanamurthy.hk@gmail.com

## ABSTRACT

*Data mining is the practice to search large amount of data to discover data patterns. Data mining uses mathematical algorithms to group the data and evaluate the future events. Association rule is a research area in the field of knowledge discovery. Many data mining researchers had improved upon the quality of association rule for business development by incorporating influential factors like utility, number of items sold and for the mining of association data patterns. In this paper, we propose an efficient algorithm to find maximal frequent itemset first. Most of the association rule algorithms used to find minimal frequent item first, then with the help of minimal frequent itemsets derive the maximal frequent itemsets, these methods consume more time to find maximal frequent itemsets. To overcome this problem, we propose a new approach to find maximal frequent itemset directly using the concepts of subsets. The proposed method is found to be efficient in finding maximal frequent itemsets.*

## 1. INTRODUCTION

Data mining is a synonym for another popular term knowledge discovery in database (KDD), the KDD process is shown in fig.1 [1]. There are three processes in KDD (preprocessing, data mining, post processing), preprocessing executed before data mining techniques are applied to the data. The preprocessing process includes data cleaning, data integration, data selection and data transformation. The main process of KDD is the data mining process, in this process different algorithm are applied to produce hidden knowledge. After, another process called post processing, evaluates the mining result according to user's requirements and domain knowledge. Regarding the evaluation results, the knowledge can be presented if the result is satisfactory, otherwise we have to run few or all of those processes again until we get the satisfactory result.

In KDD process, initially clean and integrate the databases and data source may come from different databases in which may have some inconsistences and redundancy. Clean the data source by removing noises or make some compromises. After clean and integration of databases, select related data from the integrated resources and transform them into a format that is ready to be mined.

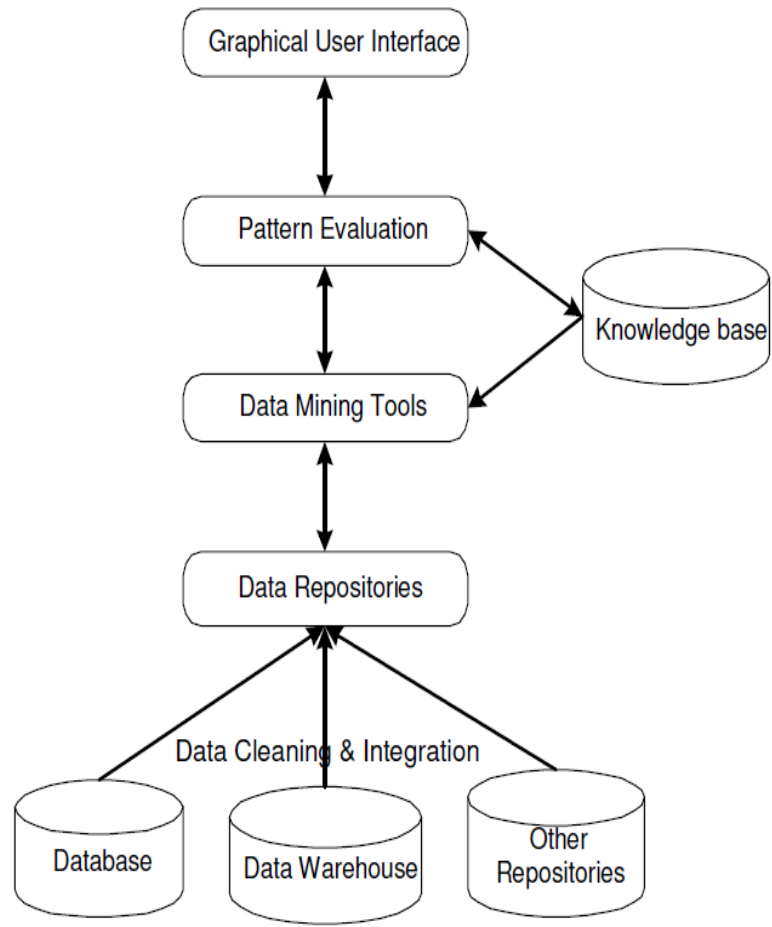

Figure 1: Knowledge Discovery in Database processes.

With the popularization of computer and development of Database Technology, more and more data are stored in large databases. Obviously, it is impossible to find useful information without using efficient methods. Data Mining (DM)[2] techniques have emerged as a reflection of this request. Association rules mining, an important research direction aims to find out the dependence among multiple domains based on a given degree of support and credibility.

Association rules mining process is divided into two steps. The first step is to find the frequent item-sets whose support degree is larger than the initial support degree from the transaction database; the second step is to generate the rules of value from the frequent item-sets, and the acquisition of frequent item-sets is the key step during mining association rules procedure. In 1993, R. Agrawal first promoted an association rule mining algorithm named Apriori algorithm[3].This algorithm's basic idea is to identify all the frequent sets whose support is greater than minimum support. Frequent item generates strong association rule, which must satisfy minimum support and minimum confidence. An Apriori idea is a brief description of the core algorithm is that has two key steps: the connecting step and the pruning step [4].

- Connecting step: In order to identify the L(k) (a frequent k set), a candidate k-items (C(k)) can be generated by L(k)-1 and its connections, which elements of L(k)-1 can be connected.

- Pruning step: C(k) is a superset of L(k) whose members may or may not be frequent, but all the frequent sets are included in C(k). If scanning database, each count of a candidate in C(k) can be determined, also L(k) (frequent candidates whose count is not less than the minimum support count). However, C(k) may be large, its calculation amount also be lots. For compression of C(k) the Apriori may be used: any non-frequent (k-1) items can not be subsets of frequent k-items. Therefore, if (k-1) items of a candidate k-items is not in L(k), then the candidate cannot be frequent, which can be deleted from C(k).

Subsequent researchers have given a lot of improvement for the AA. However, all of these improved algorithms have the following problems in varying degrees. The first problem is that algorithms need more time complexity to produce the candidate frequent item-sets. And the second is that algorithms have to scan the transaction database many times to do the pattern-matching for candidate frequent item-sets. These two issues are both the hotspots and difficulties during current research on mining association rules. In our paper, we promote a faster and more efficient algorithm based on the classical AA.

## 2. BASIC CONCEPTS

Data Mining is a method that extracts some kind of information knowledge which cannot be discovered easily, but contains certain regularity from the massive primary data [5]. Let I be a set of items and D a database of transactions. Every transaction is a set of distinct items (itemset) from I. An itemset with k items is referred to as a k-itemset. The support of an itemset X, denoted as $\sigma(X)$, is the total number of transactions in which that itemset occurs as a subset. A second formal definition for the support of an itemset X is given by Agrawal. An itemset X has a support of s if s% of transactions in D contains X as a subset. This second formal definition is somewhat more rigorous, as it emphasizes that the maximum support of an itemset cannot exceed the total number of transactions in D. An itemset is called frequent if its support is greater than a user-defined minimum support value. A frequent k-itemset X is maximal if no other k'-itemset (where k < k') contains X as a subset.

An association rule is an expression $X \Rightarrow Y$, where X and Y are disjoint itemsets. An important note is that an association rule must not be considered not only as an implication, but rather as a coexistence of the two itemsets and support is given by the support of the $X \cup Y$ itemset. The confidence of an association rule is the conditional probability that a transaction contains Y, given that it contains X, and computed using the formula $c(X \Rightarrow Y) = \sigma (X \cup Y) / \sigma(X)$. Minimum confidence of a rule is a user defined value. An association rule is strong if it has a support greater than minimum support value and confidence greater than the minimum confidence value.

## 3. THEORETICAL BACKGROUND

**Association rules:** Association rules of the form $\{X1, X2 \ldots . Xn\} \rightarrow Y$, meaning that if we find X1,X2….Xn in the market basket, then we have a good chance of finding Y. We normally would search only for rules that has confidence above the threshold and may also ask that the confidence be significantly higher, than it would be if items were placed at random into baskets.

**Frequent itemsets:** An itemset whose support is greater than or equal to a minimum support threshold is known as frequent itemsets. In many situations, we only care about association rules involving sets of items that appear frequently in baskets. For example, we cannot run a good marketing strategy involving items that no one can buys, thus data mining starts with the assumption that we only care about sets of items with high support; i.e., they appear together in many baskets. Then find the association rules only involving a high-support set of items i.e, {X1, X2,X3. . .Xn ,Y } must appear in at least a certain percent of the baskets, called the support threshold.

What is the use of studying association rules?

- With the development of e-commerce and logistics, online shopping plays an increasingly important role in people's life. Some well-known e-commerce site gets lots of benefits from mining association rules. These online shopping sites use mining association rules to get useful information from the huge database, and then set the commodity in a bundle that the customer intends to purchase together. And there are also some shopping sites which use them to set the appropriate cross-selling, where the customer who bought one product will see other related commodities advertised [6].
- In Amazon.com they used association mining to recommend you the items based on the current item you are browsing/purchasing.
- Another application is the Search engines where after you type in a word, it searches for frequently associated words that the user types after that particular word. [7]

## 4. RELATED WORK

There have been a number of attempts to find to maximal frequent itemsets. In [8], they used hash-based method to discover maximal frequent set (HMFS), the HMFS method combines DHP (Direct Hushing arid Pruning) and the Pincer-Search algorithms. In [9], a method called smartminer gathers and passes tail information and uses a heuristic function which uses the tail information to select the next node. Smartminer generates a smaller search tree requires a smaller number of supports counting and does not require superset checking. In [10], a new algorithm called data stream mining for maximal frequent itemsets (DSM-MFI), which mines the set of all maximal frequent itemsets in windows over data streams. A new summary data structure called summary frequent itemset forest (abbreviated as SFI forest) is for incremental maintaining the essential information about maximal frequent itemsets embedded in the stream. In [11], frequent pattern list (FPL) and bit string technique deals a novel algorithm for mining maximal frequent itemsets based on frequent pattern list (FPLMFI-Mining). FPLMFI-mining conducts various operations on FPL and it utilizes bit string and-operation to test maximal frequent itemsets. In [12], an algorithm based on a frequent pattern graph to find maximal frequent itemsets, breadth-first-search and depth-first-search techniques are used to produce all maximal frequent itemsets of a database.

Researchers had given a lot of improvement for the Apriori algorithm. However, all of these improved algorithms have the following problems in varying degrees. The first problem is that algorithms need more time complexity to produce the candidate frequent item-sets. And the second is that algorithms have to scan the transaction database many times to do the pattern-matching for candidate frequent item-sets. The proposed method is efficient to find maximal frequent itemset if frequent itemsets present at the initial stage.

## 5. PROPOSED METHOD

Fig.2 shows activity diagram of MFIF(proposed method) method to find maximal frequent item first. Instead of finding minimal frequent itemset first, we developed a new efficient method to find maximal frequent itemset first.

**Procedure:**

**Step1:** Count the number of items present in each transaction and put in an array a[ ].
**Step2:** Find the transactions having maximum items (value) in the array a[ ].
**Step3:** If Count (max (a[ ]) ) ≥ min_sup then transfer those transactions to an another array arr[ ][ ],else find subsets.
**Step4:** Compare each transaction in arr[ ][ ] with other transactions.
**Step5:** Take a Counter C and increase the counter if we found similar itemsets in arr[ ][ ].
**Step6:** If {C ≥ min_sup} then itemset will be the most frequent itemset.
**Step7:** if C<min_sup then find the subsets of all transactions and store it in an array sub[ ][ ].
**Step8:** max = max-1.
**Step9:** add the transactions of sub[ ][ ] to arr[ ][ ].
**Step10:** Repeat from **step3** until frequent itemset is found.

### Procedure explained with the example:

Here we took 5 transactions and 3 items I1,I2 and I3 to find maximal frequent itemset shown in below table.

| TRANSACTIONS(T) | I1 | I2 | I3 |
|---|---|---|---|
| 1 | 1 | 1 | 0 |
| 2 | 1 | 0 | 1 |
| 3 | 1 | 1 | 1 |
| 4 | 1 | 0 | 1 |
| 5 | 1 | 0 | 0 |

**Step1:** Count the number of items present in each transaction, i.e. in T1, T2, T3, T4 and T5, and store it an array A[ ] shown in below table.

| COUNT OF ITEMS IN EACH TRANSACTION (A[ ]) |
|---|
| 2 |
| 2 |
| 3 |
| 2 |
| 1 |

**Step2:** Find the transaction having maximum items (value) in the array A[ ], transaction T3 have maximum value i.e. 3 .shown in below.

| TRANSACTIONS(T) | I1 | I2 | I3 |
|---|---|---|---|
| 3 | 1 | 1 | 1 |

**Step3:** If Count (max (A[ ]) ) ≥ min_sup then transfer those transactions to an another array arr[ ][ ], else find subsets. I.e. number of transactions having count value 3 is 1, for the 60% support, minimum support is 2. Count value does not meet the minimum support {Count (max (a[ ]) )<min_sup , i.e.(1<2) }. So find the subsets of T3 shown in below table.

If Count (max (a[ ]) ) ≥ min_sup (NOT IN THE ABOVE CASE)

{

Then transfer those transactions to an another array arr[ ][ ],

**Step4:** Compare each transaction in arr[ ][ ] with other transactions.

**Step5:** Take a Counter C and increase the counter if we found similar itemset in arr[ ][ ].

**Step6:** If {C ≥ min_sup} then itemset will be the most frequent itemset.

**Step7:** if C<min_sup then find the subsets of all transactions and store it in an array sub[ ][ ] shown in below table.

}

| SUBSETS OF TRANSACTION 3 | | |
|---|---|---|
| 0 | 1 | 1 |
| 1 | 0 | 1 |
| 1 | 1 | 0 |

**Step8:** max = max-1, I.e. max value is 3, max-1 is 2. Find the transactions that count value is 2 i.e. T1, T2 and T4 shown in below table.

| TRANSACTIONS HAVE COUNT VALUE 2 | | | |
|---|---|---|---|
| **1** | 1 | 1 | 0 |
| **2** | 1 | 0 | 1 |
| **4** | 1 | 0 | 1 |

**Step9:** Combine the transactions i.e. subsets of transaction-3 and transactions have count value 2 shown in below table.

| { SUBSETS OF TRANSACTION 3 } + { TRANSACTIONS HAVE COUNT VALUE 2 } | | |
|---|---|---|
| **I1** | **I2** | **I3** |
| 0 | 1 | 1 |
| 1 | 0 | 1 |
| 1 | 1 | 0 |
| 1 | 1 | 0 |
| 1 | 0 | 1 |
| 1 | 0 | 1 |

**Step10:** Compare each transaction with other transactions and increment the count if the itemsets are same in the transaction. If count value is greater than minimum support, than that itemset will be the most frequent itemset. In the above table item I1 and I3 appearing 3 times, it meets the minimum support. So I1 I3 is the maximal frequent itemset and its subsets I1 and I3 are always frequent.

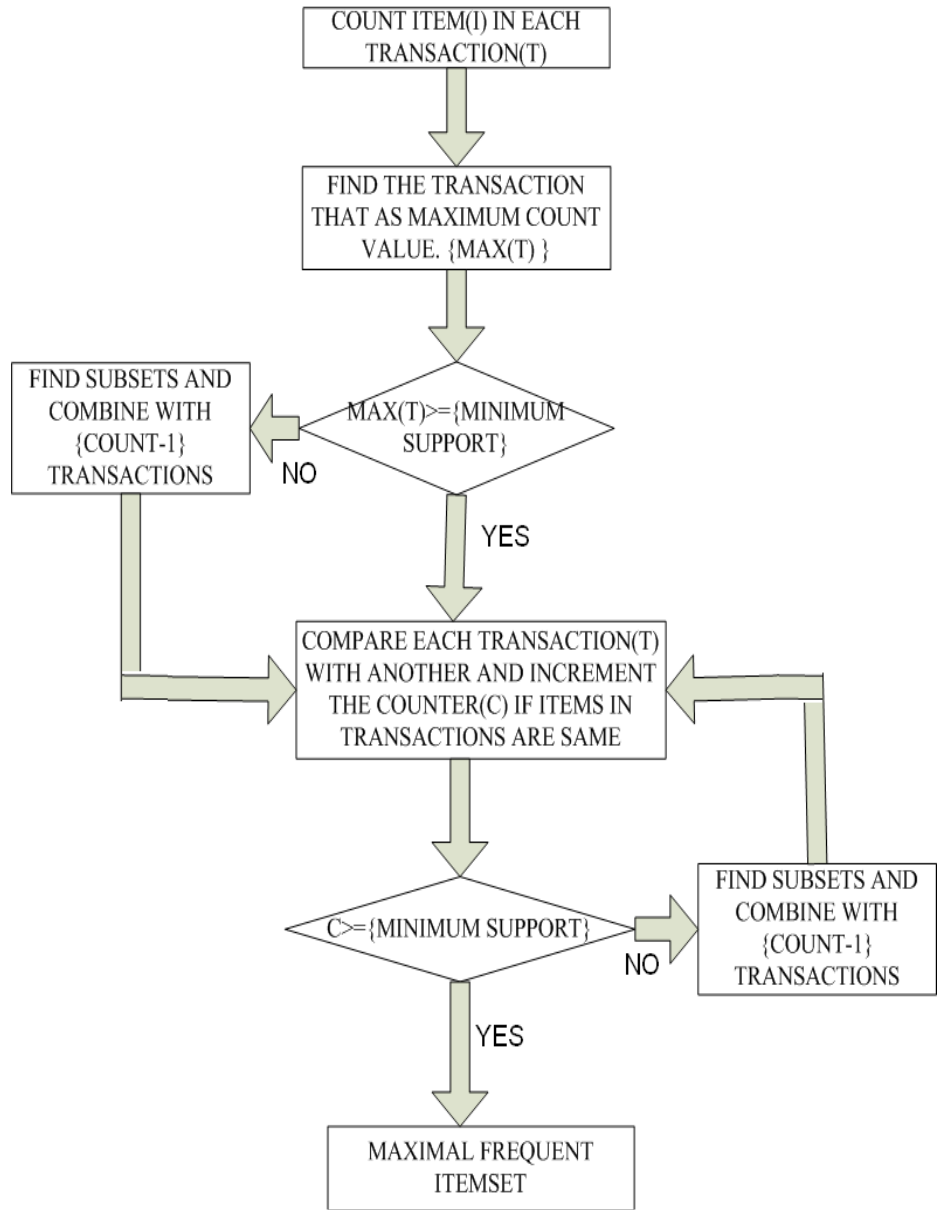

Figure 2: Procedure to find maximal frequent itemsets.

In this section we are presenting the proposed method to find maximal frequent itemsets. The working procedure is divided into 2 algorithms. The first is MFIF algorithm and the second is SUBSET FORMATION algorithm.

## MFIF ALGORITHM

**Precondition::** i=0, max=0, Count=0;
a [ ]← Count ( I ) in each T;
// I represents Items //
For (1 to n transaction)
If( a[i]>max)
max=a[i]
End if
End for

MOVE:
If count(max) ≥ min_sup
move max itemset to new_arr[ ][ ]
End if
FIND:
For all transaction in new_arr[ ] [ ]
Compare each I with (I-1) Items;
If ( I=(I-1) )
Count++;
End if
End for
If (Count ≥ min_sup)
$L_i$ ←All Itemsets with min_sup;
Else
max=max-1;
End if
Create subsets of all transactions in
new_arr[ ][ ] and store in sub_arr[ ][ ].
goto MOVE;
fin_arr[ ][ ]=new_arr[ ][ ]+sub_arr[ ][ ];
goto  FIND;

## SUBSET FORMATION ALGORITHM

For all transaction in new_arr[i][j]
temp[k]=j;
k++;
End for

Initialize l=l+i*4
For all l less than or equal to k+i*4
l++;
End for
For all m items
item[l][m]=new_arr[u][v];
item[l][temp[w]]=0;
w++;
v=0;
End for
Repeat until all the subsets are formed.

## 6. EXPERIMENTAL ANALYSIS

Fig.3 and fig.4 shows the results of proposed method. The proposed method takes less time to find maximal frequent itemset. Fig.3 consists of 10 transactions of 20 items as input, in which two transactions have 12 items and the values are similar; it meets 20 percent of minimum support, hence 12 itemset results as most maximal frequent itemset. The experiment is done till 10000 transactions.

Another example, fig.4 consists of 10 transactions of 20 items as input, here only one transaction has 13 items, and count of 13 itemset transactions will be 1, which does not meet the minimum support. So subset formation is done. Subsets will be generated of 12 items from the transaction of 13 itemset, later the generated subsets and other transactions which have 12 items will move and combine in one array and compare the subsets .If count value is greater than minimum support, then that set will be the Maximal frequent itemset.

Figure.3 MFIF Result: 12 itemset resulted as maximal frequent itemset.

Figure.4 MFIF Result: 12 itemset resulted as maximal frequent itemset with subset generation.

35

Complexity of the Apriori algorithm depends on the number of itemsets present in the transaction, i.e. if transaction has 'n' items, then we have to consider the items starting from 1 frequent itemset till we find out the 'n' frequent itemsets, so complexity increases as 'n' value increases.

MFIF(proposed algorithm) results in less time complexity compared to Apriori; when the itemsets are large, it does not depends on the value 'n'. Complexity increases only at the generation of subsets of each itemsets, and yields less time complexity if maximal frequent itemset found at the initial stage.

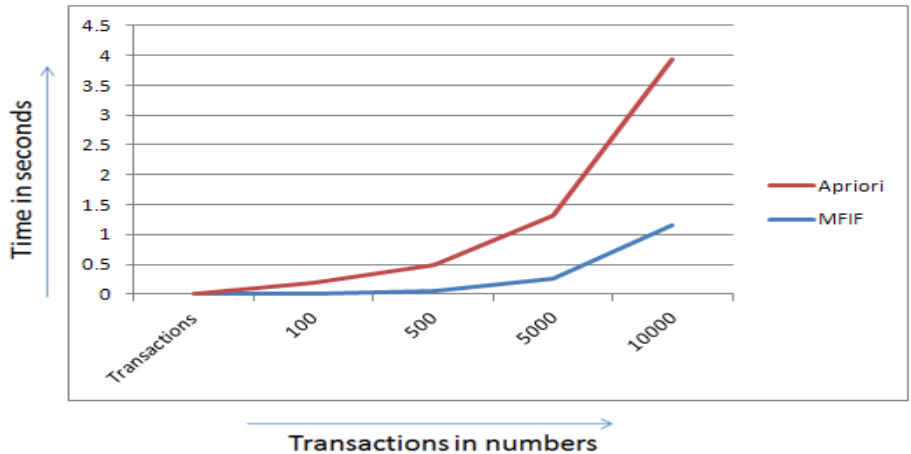

Figure.5. Graphical representation of time Complexity Comparison between Apriori and MFIF for frequent 12 itemsets.

| Transactions | MFIF (time in seconds) | Apriori (time in seconds) |
|---|---|---|
| 100 | 0.016 | 0.187 |
| 500 | 0.062 | 0.422 |
| 5000 | 0.266 | 1.047 |
| 10000 | 1.156 | 2.781 |

Table 1: Comparison between Apriori and MFIF

The results are shown in Table 1 and graphical comparison in fig.5. Time taken by MFIF and Apriori for 100, 500, 5000, 10000 transactions is shown in the table. Time complexity of MFIF is less than Apriori. Complexity of Apriori will increases as the number of items in the frequent itemset increases and MFIF complexity does not depend on the number of itemsets present. But the time complexity increases only at the time of subset generation.

The results shows that, Apriori takes more time because it takes 12 scans to find out 12 element frequent set and MFIF takes 2 scans.

# 7. ADVANTAGES

The main advantages of MFIF method are as follows:

- Too much memory space is not required for generation of subsets, because at a time only one level of element subsets are generated; as shown above only 12 element subsets are generated.
- Any element frequent set can be got in a single scan by subset creation method, which will help in applying any search method to traverse and get maximal frequent itemset, and it helps in reducing the scans drastically.
- Pruning is not required as it is done in Apriori algorithm. In MFIF maximal frequent itemset is got at first and its subsequent subsets will be frequent. So, MFIF provides an effective representation of frequent itemset.

# 8. CONCLUSION

In data mining, association rule learning is a popular and well researched method for discovering interesting relations between objects in large databases. An efficient way to discover the maximal frequent set can be very important in some kinds of data mining problems .The maximal frequent set provides an effective representation of all the frequent itemsets. Discovering maximal frequent itemsets implies immediate discovery of all frequent itemsets. This paper presents a new algorithm that can efficiently discover the maximal frequent set. The top-down searching strategy is adopted in this algorithm. This approach can be very significant and effective to find maximal frequent itemset.

## REFERENCES

[1] Qiankun Zhao and Sourav S. Bhowmick. Association Rule Mining: A Survey. Nanyang Technological University, Singapore.

[2] Arun K Pujari. Data mining concepts and techniques. Universities Press, 2001.

[3] R. Agrawal and R. Srikant. Fast algorithms for mining association rules in large databases. In Proceedings of the 20th international conference on Very Large Data Bases (VLDB'94), pages 478-499. Morgan Kaufmann, September 1994.

[4] Chengyu and Xiong Ying. Research and improvement of Apriori algorithm for association rules. In Intelligent Systems and Applications (ISA), 2010 2nd International Workshop on, pages 1 -4, may 2010.

[5] Wei Yong-Qing, Yang Ren-Hua, and Liu Pei-Yu. An improved Apriori algorithm for association rules of mining. In IT in Medicine Education, 2009. ITIME '09. IEEE International Symposium on, volume 1, pages 942 -946, aug. 2009.

[6] Guo Hongli and Li Juntao. The application of mining association rules in online shopping. In Computational Intelligence and   Design (ISCID), 2011 Fourth International Symposium on, volume 2, pages 208 -210, oct. 2011.

[7] Lu Nan, Zhou Chun-Guang, and Cui Lai-Zhong. The application of association rules algorithm on web search engine. In Computational Intelligence and Security, 2009. CIS '09. International Conference on, volume 2, pages 102 -108, dec. 2009.

[8] Don-Lin Yang, Ching-Ting Pan and Yeh-Ching Chung , "An efficient hash-based method for discovering the maximal frequent set," Computer Software and Applications Conference, 2001. COMPSAC 2001. 25th Annual International, vol., no., pp.511-516, 2001

[9] Qinghua Zou, Wesley W. Chu and Baojing Lu, "SmartMiner: A depth first algorithm guided by tail information for mining maximal frequent itemsets" Proceedings 2002 IEEE International Conference on Data Mining. ICDM 2002 , vol., no., pp.570 – 577, 2002

[10] Hua-Fu Li, Suh-Yin Lee and Man-Kwan Shan, "Online mining (recently) maximal frequent itemsets over data streams", Research Issues in Data Engineering: Stream Data Mining and Applications, 2005. RIDE-SDMA 2005. 15th International Workshop on , vol., no., pp. 11- 18, 3-4 April 2005

[11] Jin Qian and Feiyue Ye, "Mining maximal frequent itemsets with frequent pattern list", Fuzzy Systems and Knowledge Discovery, 2007. FSKD 2007. Fourth International Conference on, vol.1, no., pp.628-632, 24-27 Aug. 2007

[12] Bo Liu and Jiuhui Pan, "A graph based algorithm for mining maximal frequent itemsets", Fuzzy Systems and Knowledge Discovery, 2007. FSKD 2007. Fourth International Conference on, vol.3, no., pp.263-267, 24-27 Aug. 2007